\documentclass[conference, final, letterpaper]{IEEEtran}

\usepackage{hyperref}
\usepackage{ifpdf}
\usepackage{cite}
\ifpdf
    \usepackage[pdftex]{graphicx}
    \graphicspath{{./fig/}}
    \DeclareGraphicsExtensions{.pdf}
\else
    \usepackage[dvips]{graphicx}
    \graphicspath{{./fig/}}
    \DeclareGraphicsExtensions{.eps}
\fi
\usepackage{theorem}
\usepackage[cmex10]{amsmath}
\usepackage{amssymb}
\usepackage{stmaryrd}

\newcommand{\erasure}{\ensuremath{\vartimes}}
\newcommand{\dmin}{\ensuremath{d_\mathrm{min}}}
\newcommand{\F}{\ensuremath{\mathbb{F}}}
\newcommand{\R}{\ensuremath{\mathbb{R}}}
\renewcommand{\vec}[1]{\ensuremath{\mathbf{#1}}}
\newcommand{\hs}{\ensuremath{h_\sigma}}
\newcommand{\ez}{\ensuremath{\varepsilon_0(\tau)}}
\newcommand{\tss}{\ensuremath{{\tau_\sigma^\star}}}
\newcommand{\EbNO}{\ensuremath{E_b/N_0}}

\newcommand{\coloneq}{\ensuremath{\mathrel{\mathop:}=}}
\newcommand{\eqcolon}{\ensuremath{=\mathrel{\mathop:}}}

\newcommand{\RS}{\ensuremath{\mathcal{RS}}}
\newcommand{\A}{\ensuremath{\mathcal{A}}}
\newcommand{\B}{\ensuremath{\mathcal{B}}}

{\theoremstyle{plain}
   \newtheorem{problem}{Problem}}

{\theoremstyle{plain}
   \newtheorem{proposition}{Proposition}}

\begin{document}

\title{Adaptive Single--Trial Error/Erasure Decoding of Reed--Solomon Codes}

% Allow \thanks despite in conference mode
\IEEEoverridecommandlockouts

\author{
\authorblockN{Christian Senger, Vladimir R. Sidorenko, Steffen Schober, Martin
Bossert}\thanks{Parts of this work have been presented at the recent results poster session of IEEE
ISIT 2010. The authors have been supported by DFG,
Germany, under grants BO~867/22-1 and BO~867/21-1. Vladimir
Sidorenko is on leave from IITP, Russian Academy of Sciences, Moscow, Russia.}
\authorblockA{\small Inst. of Telecommunications and Applied Information
Theory\\
Ulm University, Ulm, Germany \\
\{
christian.senger$\;\vert\;$vladimir.sidorenko$\;\vert\;$steffen.schober$\;\vert\;$martin.bossert\}
@uni-ulm.de}
\and
\authorblockN{Victor V. Zyablov}
\authorblockA{\small Inst. for Information Transmission Problems\\
Russian Academy of Sciences, Moscow, Russia \\
zyablov@iitp.ru}
}

\maketitle

\begin{abstract}
Algebraic decoding algorithms are commonly applied for the decoding of
Reed--Solomon codes. Their main advantages are low computational complexity and
predictable decoding capabilities. Many algorithms can be extended for
correction of both errors and erasures. This enables the decoder to exploit
binary quantized reliability information obtained from the transmission channel:
Received symbols with high reliability are forwarded to the decoding algorithm
while symbols with low reliability are erased. In this paper we investigate
adaptive single--trial error/erasure decoding of Reed--Solomon codes, i.e. we
derive an adaptive erasing strategy which minimizes the residual codeword error
probability after decoding. Our result is applicable to any error/erasure
decoding algorithm as long as its decoding capabilities can be expressed by a decoder
capability function. Examples are Bounded Minimum Distance decoding with the Berlekamp--Massey- or the Sugiyama algorithms and the Guruswami--Sudan list decoder.
\end{abstract}

% ##############################################################################
\section{Introduction}\label{sec:intro}

Using algebraic error/erasure decoders for pseudo--soft decoding of {\em
Reed--Solomon (RS)} codes dates back to Forney \cite{forney:1966b,
forney:1966a}. His {\em Generalized Minimum Distance (GMD)} decoding applies an
error/erasure decoder multiple times, each time with an increased number of erased
most unreliable symbols of the received word. In very good channels, the residual codeword error
probability of GMD decoding approaches that of {\em Maximum Likelihood (ML)}
decoding if the number $z$ of such decoding trials is sufficiently large, i.e.
$z\approx \dmin/2$, where $\dmin$ is the minimum {\em Hamming distance} of the RS
code. Thus, the computational complexity of GMD decoding is $\dmin/2$--times
that of errors--only decoding. Roughly, we can say $\dmin\in\mathcal{O}(n)$, which
means that quadratic decoding complexity in the code length $n$ becomes cubic and so on.
K\"otter \cite{koetter:1993} provided a modification of the
Berlekamp--Massey algorithm that essentially computes all decoding trials at
once, i.e. without increasing complexity. A similar result was achieved in
\cite{kampf_bossert:2010a, kampf_bossert:2010b} for the Euclidean algorithm and in
\cite{sorger:1993} using Newton interpolation. Up to our knowledge, there is no such modification of
{\em Guruswami--Sudan (GS)} \cite{guruswami_sudan:1999} list decoding -- which can be considered as state of the art
in algebraic decoding -- so far. However, K\"otter and Vardy provided a modification of the GS algorithm that is capable of exploiting soft information \cite{koetter_vardy:2003}.

GMD decoding is per se a {\em fixed} approach, i.e. the erasing strategy is constant for each
received word. {\em Adaptive} variants
have been investigated in \cite{kovalev:1986, sidorenko_senger_bossert_zyablov:2008,
sidorenko_chaaban_senger_bossert:2009}. The respective authors show that the number $z$ of decoding
trials can be reduced significantly, if the erasing strategy is optimally calculated for every
single received word. The aforementioned papers focus on the maximization of the achievable
{\em decoding radius}, i.e. the maximum correctable number of errors in the received word. In contrast to that, our objective is to minimize the residual codeword error
probability after decoding. We achieve this using a technique first introduced in 2010
\cite{senger_sidorenko_schober_bossert_zyablov:2010} for optimal error/erasure decoding of binary
codes. As in the latter paper, we restrict ourselves to one
single decoding trial, i.e. $z=1$, for simplicity.

The rest of the paper is organized as follows. In Section~\ref{sec:ee} we give an overview of
error/erasure decoding and introduce the required notations. We introduce the decoder
capability function which allows to derive the optimal erasing strategy in a general manner.
Here and in the following, {\em optimal} means {\em minimizing the residual codeword error
probability}. In Section~\ref{sec:strategy}, we derive an optimal adaptive erasing strategy for one
single decoding trial. Section~\ref{sec:approx} describes two computationally efficient
approximations of the optimal strategy, one of them with complexity quadratic in the code length
$n$. In Section~\ref{sec:sim} we show the potential of single--trial error/erasure
decoding in terms of achievable residual codeword error probability as well as the quality of the
two approximative variants by simulation. The paper is wrapped up with conclusions and an outlook in
Section~\ref{sec:conc}.

% ##############################################################################
\section{Reliability--Based Error/Erasure Decoding and Decoder Capability Functions}\label{sec:ee}

Consider the RS code $\RS(q; n, k, \dmin)$ of length $n$, dimension $k$ and minimum distance
$\dmin$ over the extension field $\F_q$. Thereby, $q\coloneq p^m$ for some prime number $p$ and some
integer $m$. The transmitter encodes an information vector $\vec{a}\in\F_q^k$ into a codeword
$\vec{c}\coloneq (c_0, \ldots, c_{n-1})\in\RS\subseteq\F_q^n$. Each symbol $c_i$, $i=0, \ldots, n-1$, is
mapped to a symbol $x_i$ of a modulation alphabet $\A\subseteq\R^2$ resulting in
the modulated codeword $\vec{x}\coloneq (x_0, \ldots, x_{n-1})\in\A^n$. This modulated codeword is
transmitted over the channel, where it is distorted by two--dimensional {\em Additive White
Gaussian Noise (AWGN)}. At the receiver, the received word
$\vec{x}+\vec{e}=(y_0, \ldots, y_{n-1})\eqcolon \vec{y}\in\left(\R^2\right)^n$ is obtained. It is the sum of the modulated
codeword and an error word $(e_0, \ldots, e_{n-1})\eqcolon \vec{e}\in\left(\R^2\right)^n$. Each symbol $y_i$ is mapped to the
closest (in Euclidean metric) modulation point of $\A$, the result of this procedure is the {\em
hard decision} $\vec{\tilde{x}}\coloneq (\tilde{x}_0, \ldots, \tilde{x}_{n-1})\in\A^n$ of $\vec{y}$. This
hard decision can be fed into the inverse mapper function to obtain a received vector
$\vec{r}\coloneq (r_0, \ldots, r_{n-1})\in\F_q^n$, which in turn can be fed into any algebraic hard decision decoder for
$\RS$.

The hard decision, i.e. the mapping from received symbols $y_i$ to closest modulation points
$\tilde{x}_i$, is error--prone since it is not necessarily correct: The received $y_i=x_i+e_i$ might be closest in Euclidean metric to an $x_i'\neq x_i$. We refer to the probability of
an incorrect hard decision as {\em unreliability} of a received symbol and denote it by
$\hs(\tilde{x}_i)$. Here,
\begin{equation*}
\sigma \coloneq \sqrt{\frac{1}{\log_2{\mid\A\mid}}\cdot\frac{n}{k}\cdot\frac{10^{-\frac{\mathrm{\EbNO}}{10}}}{2}}
\end{equation*}
is the AWGN standard deviation for $\EbNO$ given in dB. Note that due to the one--to--one mapping between $\tilde{x}_i$ and $r_i$ we can
as well write the unreliability as a function of the de--modulated received symbols, i.e. $\hs(r_i)\coloneq \hs(\tilde{x}_i)$.
By definition, we have
\begin{align}
  \hs(r_i)  &= 1-\Pr(\tilde{x}_i\;\text{transmitted}\mid y\;\text{received})\nonumber\\
          &=  1-\frac{%
              \Pr(y\mid \tilde{x}_i)\Pr(\tilde{x}_i)
              }{%
              \Pr(y)
              }\nonumber\\
          &= 1-\frac{%
              \Pr(y\mid \tilde{x}_i)\Pr(\tilde{x}_i)
              }{%
              \sum_{x\in\A} \Pr(y\mid x)\Pr(x)
              }\nonumber\\
          &= 1-\frac{%
              \Pr(y\mid \tilde{x}_i)
              }{%
              \sum_{x\in\A} \Pr(y\mid x)
              },\label{eqn:unrel_exact}
\end{align}
where the last equality follows from the assumption of equiprobable codeword symbols. In practice, the calculation of (\ref{eqn:unrel_exact}) is not feasible for large modulation alphabets $\A$, hence we use the {\em nearest neighbor approximation}
\begin{equation}\label{eqn:unrel_approx}
  \hs(r_i)\approx 1-\frac{%
              \Pr(y\mid \tilde{x}_i)
              }{%
              \sum_{x\in\B(\tilde{x}_i)} \Pr(y\mid x)
              },
\end{equation}
where $\B(\tilde{x}_i)\subseteq\A$ is the set of nearest neighbors of $\tilde{x}_i$. This allows to store all possible values of $\hs(r_i)$ in a comparatively small lookup table as the following example demonstrates.

\begin{figure}[htbp]
\centering
\includegraphics[width=150pt]{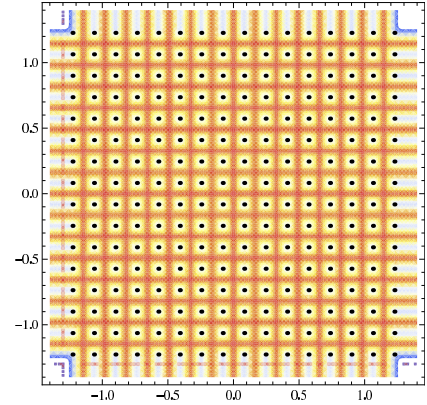}
\caption{Unreliability $\hs(r_i)$ for the Euclidean plane, $256$--QAM, $8$--bit--quantization, AWGN@$18$ dB.}
\label{fig:unrel_full}
\end{figure}

Assume {\em $256$--Quadrature Amplitude Modulation (QAM)} with average signal energy one, Gray mapping and $8$--bit--quantization. Without nearest neighbor approximation, this leads to a lookup table with $65536/4=16348$ entries, each containing two integers for the coordinates and one real number for the unreliability value $\hs(r_i)$. Considering only the nearest neighbors, the lookup table consists of only $64+128+128=320$ such entries if symmetries within the QAM decision regions are exploited. This allows to store lookup tables for many different $\EbNO$ values, e.g. up to the precision of the (required) channel estimation. Fig.~\ref{fig:unrel_full} shows a density plot of $\hs(r_i)$ for the complete Euclidean plane, the black dots mark the modulation points, darker color marks higher unreliability. Note that the unreliabilities for most decision regions are either rotated versions of each other or they coincide. Furthermore, the unreliabilities of the decision regions are either point symmetric to their modulation point (regions in the center) or symmetric to a line through the modulation point (border and corner regions). This allows to discard $3/4$ of the quantization intervals in the first case and $1/2$ of the intervals in the second case.

At the receiver, the unreliability is calculated (or taken from the lookup table) for every received symbol $r_i$, $i=0, \ldots, n-1$. W.l.o.g. we assume here and in the following that the received word $\vec{r}$ is ordered according to its symbol's unreliabilities, i.e. $\hs(r_0)\geq\cdots\geq \hs(r_{n-1})$. The idea of error/erasure decoding is to discard the $\tau$ most unreliable symbols (i.e. to erase them) since it is likely that they are erroneous. Instead of $\vec{r}$, the input vector fed into the algebraic decoder is then
\begin{equation*}
  \vec{r}_\tau=(\underbrace{\erasure, \ldots, \erasure}_{\tau\,\text{times}}, r_{\tau}, \ldots, r_{n-1}),
\end{equation*}
where the first $\tau$ symbols are replaced by the erasure marker $\erasure$. In order to do this, two conditions need to be fulfilled. First, we require an algebraic decoder which is capable of decoding both errors and erasures. Second, we must be capable of deciding how many of the most unreliable symbols should be discarded.

Algebraic error/erasure decoders for RS codes are well--known. Classical {\em Bounded Minimum Distance (BMD)} e.g. decoding using the Berlekamp--Massey- or the Sugiyama algorithms can be augmented by an erasure option \cite{blahut:1979} as can the GS list decoder \cite{guruswami_sudan:1999}. In \cite{sidorenko_schmidt_bossert:2008}, a decoder with erasure option for {\em Interleaved Reed--Solomon (IRS)} codes from \cite{schmidt_sidorenko_bossert:2006c} is applied to decode {\em $\ell$--punctured} RS codes.

The {\em decoder capability function (DCF)} of an algebraic error/erasure decoder is an inequality of the form
\begin{equation*}
  f(n, \varepsilon, \tau)>k-1,
\end{equation*}
which is true whenever the decoder can correct $\varepsilon$ errors and $\tau$ erasures in any given received word. Three important examples for error/erasure decoders and their respective $f(n, \varepsilon, \tau)$ are given in Table~\ref{tab:dcf}.

% Increase distance between tabular rows
\renewcommand{\arraystretch}{2.8}
\begin{table*}[htbp]
\begin{center}
\begin{tabular}{l|c|c}
{\bf Decoder} & $\displaystyle f(n, \varepsilon, \tau)$ & $\displaystyle \ez$\\
\hline\hline
Bounded Minimum Distance &%
$\displaystyle n-\tau-2\varepsilon$ &%
$\displaystyle \left\lceil\frac{n-k+1-\tau}{2}\right\rceil-1$
\\\hline
IRS--based (for $\ell$--punctured RS codes) &%
$\displaystyle n-\tau-\frac{\ell+1}{\ell}\varepsilon$ &%
$\displaystyle \left\lceil\frac{\ell(n-k+1-\tau)}{\ell+1}\right\rceil-1$
\\\hline
Guruswami--Sudan, $\nu\rightarrow\infty$ &%
$\displaystyle \frac{(n-\tau-\varepsilon)^2}{n-\tau}$ &%
$\displaystyle \left\lceil n-\tau-\sqrt{(n-\tau)(k-1)} \right\rceil-1$
\end{tabular}
\end{center}
\caption{Three algebraic decoders for $\RS(q; n, k, \dmin)$, their respective $f(n, \varepsilon, \tau)$--functions, and $\ez$.}
\label{tab:dcf}
\end{table*}
\renewcommand{\arraystretch}{1}

Based on the DCF of a decoder, it is straightforward to calculate the maximal number of correctable errors for any given number of erasures in a received word, i.e. the maximal number $\varepsilon$ which fulfills the DCF for fixed $\tau$. We denote this number of errors by $\ez$, where $\tau$, $0\leq\tau\leq \dmin-1$, is the number of erasures. Table~\ref{tab:dcf} shows $\ez$ for the three considered decoders.

% ##############################################################################
\section{Optimal Erasing Strategy}\label{sec:strategy}

In this section, we shall solve the following basic problem of adaptive single--trial error/erasure decoding. Thereby, our technique is similar to \cite{senger_sidorenko_schober_bossert_zyablov:2010}.

\begin{problem}\label{prob:problem1}
  For given received vector $\vec{r}\coloneq (r_0, \ldots, r_{n-1})$ with ordered unreliability values $\hs(r_0)\geq\cdots\geq \hs(r_{n-1})$ and channel state $\sigma$, find the optimal number $\tss$, $0\leq\tss\leq\dmin-1$, of erased most unreliable symbols, such that the residual codeword error probability of decoding $\vec{r}_\tss$ using a decoder with DCF $f(n, \varepsilon, \tau)$ is minimized.
\end{problem}

The foundation of our solution is basic probability theory and our aim is to express the residual codeword error probability as a function of the number $\tau$ of erased most unreliable symbols. In the following, we omit all subindices $\sigma$ for simpler reading. However, the reader should keep in mind that all functions and values depend on $\sigma$.

Given a received vector $\vec{r}$, we define the binary random variables $X_i$, $i=0, \ldots, n-1$, by
\begin{equation*}
  X_i\coloneq\left\{\begin{array}{rl}
      1, & \text{if}\;r_i\;\text{is erroneous}\,(r_i\neq c_i)\\
      0, & \text{if}\;r_i\;\text{is correct}\,(r_i= c_i)
      \end{array}\right..
\end{equation*}
By definition of $h(r_i)$, the probabilities  of the two possible outcomes of $X_i$ are $\Pr(X_i=1)=h(r_i)$ and $\Pr(X_i=0)=1-h(r_i)$. Thus, the {\em probability generating function (PGF)} of $X_i$ is
\begin{equation*}
  G_{X_i}(\rho) \coloneq 1-h(r_i)+\rho h(r_i).
\end{equation*}

After erasing $\tau$ symbols from $\vec{r}$, there are $\varepsilon$, $0\leq\varepsilon\leq n-\tau$, erroneous symbols within the non--erased $n-\tau$ symbols of $\vec{r}_\tau$. We denote their number by the discrete random variable
\begin{equation*}
  Y_\tau \coloneq \sum_{i=\tau}^{n-1} X_i
\end{equation*}
with PGF
\begin{equation*}
  G_{Y_\tau}(\rho) \coloneq \prod_{i=\tau}^{n-1} G_{X_i}(\rho).
\end{equation*}

Using the PGF of $Y_\tau$, the probability for $\varepsilon$ errors in the $n-\tau$ non--erased symbols can be calculated by
\begin{equation*}
  \Pr(Y_\tau=\varepsilon)=\left.\frac{G_{Y_\tau}^{(\varepsilon)}(\rho)}{\varepsilon!}\right|_{\rho=0},
\end{equation*}
where the superscript $^{(\varepsilon)}$ denotes the $\varepsilon$-th derivative.

Decoding $\vec{r}_\tau$ is successful if $\varepsilon\leq\ez$ and fails otherwise. Hence, we can state the residual codeword error probability after decoding $\vec{r}_\tau$ by
\begin{equation}\label{eqn:Pf}
  P(\tau)\coloneq \sum_{\varepsilon=\ez+1}^{n} \Pr(Y_\tau=\varepsilon)= 1-\sum_{\varepsilon=0}^{\ez} \Pr(Y_\tau=\varepsilon).
\end{equation}
Consequently, the optimal choice of $\tau$ (solving Problem~\ref{prob:problem1}) is
\begin{align}
  \tau^\star &\coloneq \arg\min_{0\leq\tau\leq\dmin-1}\left\{P(\tau)\right\}\nonumber\\
  &=\arg\max_{0\leq\tau\leq\dmin-1}\left\{\sum_{\varepsilon=0}^{\ez} \Pr(Y_\tau=\varepsilon)\right\}.\label{eqn:solution}
\end{align}

A closer look on the complexity of (\ref{eqn:solution}) shows that it is in $\mathcal{O}(n^3)$, see \cite{senger_sidorenko_schober_bossert_zyablov:2010} for a detailed explanation. The main task is to calculate $\Pr(Y_\tau=0), \ldots, \Pr(Y_\tau=\ez)$ for all $\tau$, $0\leq\tau\leq\dmin-1$ in order to obtain $P(\tau)$.

% ##############################################################################
\section{Computationally Efficient Erasing Strategies}\label{sec:approx}

In this section, we give two approximations of $P(\tau)$ which eventually allow to calculate approximative $\tau^\star$ with complexities in $\mathcal{O}(n^2\sqrt{n})$ and $\mathcal{O}(n^2)$, respectively.

% ------------------------------------------------------------------------------
\subsection{Approximation based on the Hoeffding Bound}\label{subsec:hoeffding}

Given a set of independent random variables fulfilling certain properties, the {\em Hoeffding bound} \cite{hoeffding:1963} states that the probability of this sum to assume values outside a small discrete interval $\mathcal{I}$  around its expectation is exponentially small. We show in \cite{senger_sidorenko_schober_bossert_zyablov:2010} that the $X_i$, $i=0, \ldots, n-1$, and their partial sum $Y_\tau$ have these properties, leading to the inequality
\begin{equation*}
 \Pr(|Y_\tau-E\{Y_\tau\}|\geq s) \leq 2 \exp\left(-\frac{s^2}{2n}\right),
\end{equation*}
where $s$ is a parameter that denotes the one--directional size $t=s/(n-\tau)$ of
\begin{equation*}
  \mathcal{I}\coloneq[E\{Y_\tau\}-t, \ldots, E\{Y_\tau\}, \ldots, E\{Y_\tau\}+t].
\end{equation*}
Setting $s>\sqrt{-\log(0.5\cdot 10^{-2}) 2n}$ guarantees that the share of the probabilities $\Pr(Y_\tau=\varepsilon)$ for $\epsilon\not\in\mathcal{I}$ in the calculation (\ref{eqn:Pf}) of $P(\tau)$ is small, cf. Table~\ref{tab:share}.

% Increase distance between tabular rows
\renewcommand{\arraystretch}{2}
\begin{table}[htbp]
\begin{center}
\begin{tabular}{c|c}
{\bf Partial Sum} & {\bf Share in $P(\tau)$}\\
\hline\hline
$\sum_{\varepsilon\in\mathcal{I}}\Pr(Y_\tau=\varepsilon)$ & $>99\%$\\
$\sum_{\varepsilon\not\in\mathcal{I}}\Pr(Y_\tau=\varepsilon)$ & $<1\%$\\
\end{tabular}
\end{center}
\caption{Percentual shares of the probabilities $\Pr(Y_\tau=\varepsilon)$ in $P(\tau)$ for $\varepsilon\in\mathcal{I}$ and $\varepsilon\not\in\mathcal{I}$.}
\label{tab:share}
\end{table}
\renewcommand{\arraystretch}{1}

As a result, all values $\varepsilon\not\in\mathcal{I}$ in  (\ref{eqn:Pf}) can be neglected and a good approximation of the residual codeword error probability is obtained by
\begin{equation*}
  P(\tau)\approx 1-\sum%
    _{\varepsilon=\max\{E\{Y_\tau\}-t, 0\}}%
    ^{\min\{E\{Y_\tau\}+t, \ez\}}%
    \Pr(Y_\tau=\varepsilon).
\end{equation*}

In \cite{senger_sidorenko_schober_bossert_zyablov:2010}, the complexity of adaptive single--trial error/erasure decoding of binary codes with the Hoeffding approximation of $P(\tau)$ is stated to be in $\mathcal{O}(n^2\sqrt{n})$, this also holds for our case of (non--binary) RS codes.

% ------------------------------------------------------------------------------
\subsection{Approximation based on $\ez$}\label{subsec:boundary}

For the second approximation, we require the following proposition. So far it is verified only by experiments and we are working on a proof.

\begin{proposition}\label{prop:unimodal}
  For fixed $\tau$, $0\leq\tau\leq\dmin-1$, $\Pr(Y_\tau=\varepsilon)$ is a unimodal function in $\varepsilon$ and its mode is determined by the expectation $E\{Y_\tau\}$.
\end{proposition}

\begin{figure}[htbp]
\centering
\includegraphics[width=252pt]{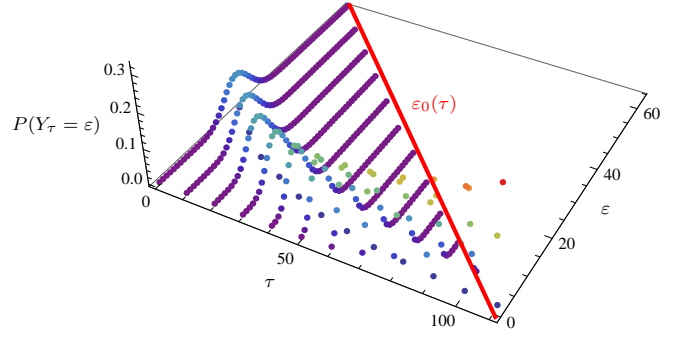}
\caption{Probabilities $\Pr(Y_\tau=\varepsilon)$ for $\RS(256; 255, 144, 112)$, AWGN@$18$ dB, $\ez$ of the GS list decoder.}
\label{fig:optimization}
\end{figure}

Fig.~\ref{fig:optimization} shows a $3$D plot of $\Pr(Y_\tau=0), \ldots, \Pr(Y_\tau=\ez)$ for all $\tau$, $0\leq\tau\leq\dmin-1$, each slice in $\varepsilon$--direction is unimodal according to Proposition~\ref{prop:unimodal} and each $E\{Y_\tau\}$ coincides with the mode of the respective $\tau$.

If the expectation is smaller than the error boundary, i.e. if $E\{Y_\tau\}\leq\ez$, then we can approximate the sum $\sum_{\varepsilon=\ez+1}^{n} \Pr(Y_\tau=\varepsilon)$ by its largest element, which is $\Pr(Y_\tau=\ez+1)$. Analogously, if $E\{Y_\tau\}>\ez$, then $\Pr(Y_\tau=\ez)$ is a good approximation of the sum $\sum_{\varepsilon=0}^{\ez} \Pr(Y_\tau=\varepsilon)$. Inserting into (\ref{eqn:Pf}) gives
\begin{equation*}
  P(\tau)\approx \left\{\begin{array}{cc}
                           1-\Pr(Y_\tau=\ez),  & \text{if}\;E\{Y_\tau\}>\ez\\
                           \Pr(Y_\tau=\ez+1), & \text{if}\;E\{Y_\tau\}\leq\ez
                          \end{array}\right..
\end{equation*}

Based on the complexity analysis of the Hoeffding approximation in \cite{senger_sidorenko_schober_bossert_zyablov:2010}, it is easy to see that the complexity of the $\ez$ approximation is in $\mathcal{O}(n^2)$. Since there are no practical decoders with lower complexity than $\mathcal{O}(n^2)$, adaptive single--trial error/erasure decoding is in the same complexity class as the error/erasure decoder itself and the computation of $\tau^\star$ increases complexity only by an additive constant.

% ##############################################################################
\section{Simulation Results}\label{sec:sim}

We investigate the potential of adaptive single--trial error/erasure decoding for the RS code  $\RS(256; 255, 144, 112)$. Two error/erasure decoders are considered: The classical BMD decoder (e.g. Berlekamp--Massey or Sugiyama) and the GS list decoder with multiplicities $\nu\rightarrow\infty$.

Performance evaluation is done by simulation and by semi--simulative upper bounds of the residual codeword error probability. For each considered $\EbNO$, we calculate an average unreliability vector $\vec{\bar{h}}\coloneq(\bar{h}_0, \ldots, \bar{h}_{n-1})$, $\bar{h}_0\geq\cdots\geq \bar{h}_{n-1}$, by averaging over $10^4$ random unreliability vectors. For each variant of $P(\tau)$ (exact, Hoeffding approximation, $\ez$ approximation), we calculate $\bar{\tau}\coloneq \tau^\star$ for $\vec{\bar{h}}$ according to (\ref{eqn:solution}). We use $\bar{\tau}$ for every received vector, which means that the simulation is in fact non-adaptive, using the optimal erasing  strategy for the average unreliability vector. The resulting residual codeword error probability curves are indeed upper bounds, it is clear that the error probability can not be higher when $\tau^\star$ is calculated for every single received vector. Precise error probabilities of errors--only decoding are obtained by inserting $\tau=0$ into (\ref{eqn:Pf}).

\begin{figure}[htbp]
\centering
\includegraphics[width=232pt]{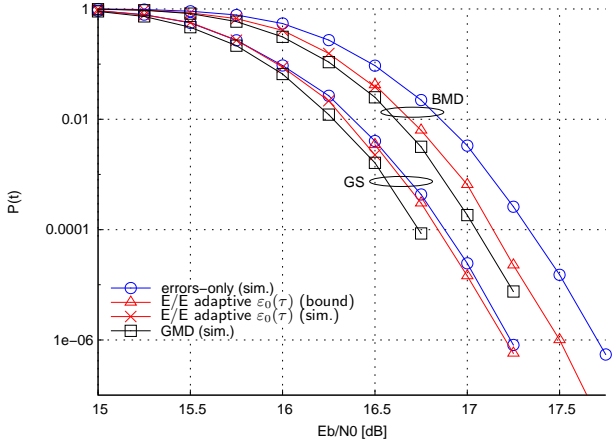}
\caption{Residual codeword error probability vs. Eb/N0 for different decoders, $\RS(256; 255, 144, 112)$. Adaptive single--trial error/erasure decoding is based on the $\ez$ approximation.\vspace{-0.5cm}}
\label{fig:sim255_144}
\end{figure}

The $\ez$ approximation is considered in Fig.~\ref{fig:sim255_144}. It shows actual simulation results for $\EbNO=15,\ldots, 16.5$ dB and the aforementioned upper bound for $\EbNO=16.5,\ldots, 17.75$ dB. Clearly, adaptive single--trial error/erasure decoding with a classical BMD decoder yields a gain of approximately $0.2$ dB for practical error probabilities. For the GS list decoder, the achievable gain is negligible. The reason for this lies in the non--linearity of the GS list decoder's $\ez$ function, whose slope gets steeper with decreasing $\tau$. This means that the benefit of transforming errors into erasures diminishes for a small number of erased symbols. The residual codeword probability curve of Forney's original $z\approx \dmin/2$--trial GMD decoding \cite{forney:1966b,forney:1966a} is given as a reference. Fig.~\ref{fig:sim255_144} shows that most of GMD's gain can be achieved by a single adaptive trial, if only the erasing strategy is chosen optimally.

\begin{figure}[htbp]
\centering
\includegraphics[width=232pt]{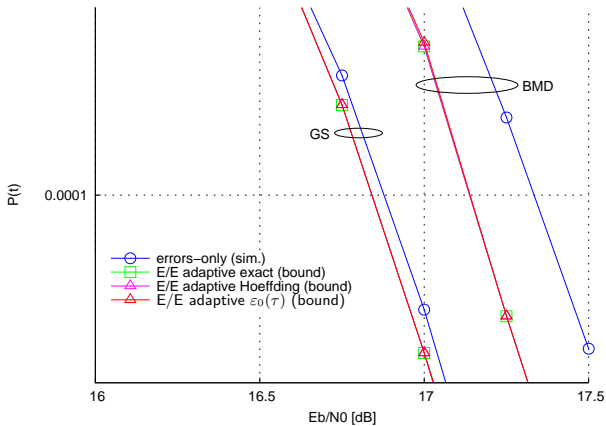}
\caption{Comparison between the exact optimal erasing strategy and the two proposed approximations, $\RS(256; 255, 144, 112)$.}
\label{fig:sim255_144_zoom}
\end{figure}

Fig.~\ref{fig:sim255_144_zoom} shows for an interesting range of residual codeword error probabilities that there is virtually no difference between exact calculation of $P(\tau)$ and the two proposed approximations. Our recommendation is to use the $\ez$ approximation whenever exact calculation of $P(\tau)$ is prohibitive. It is feasible both in terms of computational complexity ($\mathcal{O}(n^2)$) and approximation quality.

% ##############################################################################
\section{Conclusions}\label{sec:conc}

Classical error/erasure BMD decoders for RS codes are widely deployed. We presented an adaptive single--trial error/erasure decoding technique, which allows to decrease the residual codeword error probability of such decoders using a low--complexity, i.e. $\mathcal{O}(n^2)$, pre--computation step. The achievable gain of our technique is around $0.2$ dB for $\RS(256; 255, 144, 112)$. This is slightly less than the gain of GMD decoding but neither does it require a modification of the decoder itself (K\"otter's fast GMD decoder \cite{koetter:1993}) nor does it require $z\approx \dmin/2$ decoding trials (Forney's original GMD decoder \cite{forney:1966b,
forney:1966a}). Our technique is general, it can be applied to any error/erasure decoder as long as its DCF is known.

% ##############################################################################
\section*{Acknowledgments}
\addcontentsline{toc}{section}{Acknowledgments}
The authors would like to thank Dejan E. Lazich for carefully
proofreading the manuscript.

\def\noopsort#1{}

\end{document}